\documentclass[%
preprint,
 amsmath,amssymb,
 aps,
]
{revtex4-1}
\usepackage{graphicx}
\usepackage{braket}
\usepackage{bm}
\usepackage{amsmath}
\usepackage{color}
\begin{document}
\title{Triaxial deformation and the loss of the $\bm{N=28}$ shell gap}
\author{Y. Suzuki}
\affiliation{Department of Physics, Hokkaido University, 060-0810 Sapporo, Japan}
\author{M. Kimura}
\email{masaaki@nucl.sci.hokudai.ac.jp}
\affiliation{Department of Physics, Hokkaido University, Sapporo 060-0810, Japan} 
\affiliation{Nuclear Reaction Data Centre, Hokkaido University, Sapporo 060-0810, Japan} 
\affiliation{Research Center for Nuclear Physics (RCNP), Osaka University, Ibaraki 567-0047, 
Japan} 
\date{\today}

\begin{abstract}
 \noindent{\bf Background}: Recent accumulation of experimental data is revealing the nuclear
 deformation in vicinity of $^{42}{\rm Si}$. This requests systematic theoretical studies to clarify 
 more specific aspects of nuclear deformation and its causes.\\    
 \noindent{\bf Purpose}: The purpose of this study is to investigate the nature and cause of the
 nuclear deformations and its relation to the loss of the neutron magic number $N=28$ in vicinity of 
 $^{42}{\rm Si}$. \\
 \noindent {\bf Method}: The framework of antisymmetrized molecular dynamics with Gogny D1S density
 functional has been applied. The model assumes no spatial symmetry and can describe triaxial
 deformation. It also  incorporates with the configuration mixing by the generator coordinate method. \\
 \noindent {\bf Results}: We show that the shell effects and the loss of the magicity induce various
 nuclear deformations. In particular, the $N=26$ and $N=30$ isotones have triaxially deformed ground
 states. We also note that the erosion of the $N=28$ magicity gradually occurs and  has no definite
 boundaries.\\  
 \noindent{\bf Conclusion}: The present calculation predicts various nuclear deformations in
 vicinity of $^{42}{\rm Si}$, and suggests that the inter-band electric transitions are good measure
 for it. We also remark that the magicity is lost without the single-particle level inversion in the
 oblate deformed nuclei such as $^{42}{\rm Si}$.   
\end{abstract}

\maketitle

\section{introduction}
The loss of magic numbers \cite{Wilkinson1959,Talmi1960,Thibault1975,Sorlin2008}, which occurs in
unstable nuclei with neutron numbers $N=8$, 20, 28, and 50, significantly changes the structure of
Fermi surface. Hence, it is expected to considerably affect the fundamental properties of nuclei,
such as stability, size, and shape. In fact, decades of studies have shown that various phenomena
such as nuclear deformation~\cite{Warburton1990,Fukunishi1992,Motobayashi1995,Sorlin1993,Scheit1996,
Glasmacher1997,Takeuchi2012a}, shape
coexistence~\cite{Heyde2011,Otsuka2016,Nishibata2019}, radius
increase~\cite{Tanihata1985,Ozawa2001,Horiuchi2006,Nakamura2009,Takechi2012,Minomo2012}, 
and clustering~\cite{VonOertzen2006,Kimura2016} occur with 
the  loss of neutron magicity. The causes of the vanishing magic number, such as weak
bindingness~\cite{Hamamoto2012} and the effect of nuclear forces~\cite{Poves1987,Caurier2005},
especially that of tensor force~\cite{Otsuka2005,Otsuka2020}, have also been intensively studied.  

Among the neutron magic numbers, $N=28$ is the smallest one generated by the spin-orbit interaction.
Therefore, its disappearance should change the structure of the Fermi surface in a different way
from the cases of $N=8$ and 20. More specifically, the $N=28$ shell gap is composed of the neutron
$0f_{7/2}$ and $1p_{3/2}$ orbits which belong to the same major shell in the absence of spin-orbit
splitting. This means that the quenching of the $N=28$ shell gap causes quasi-degeneracy of the
orbits which have the same parity but different angular momenta by two. This will induce strong
quadrupole correlations between the nucleons near the Fermi surface, which leads to various
quadrupole deformation of the low-lying
states~\cite{Santiago-Gonzalez2011a,Utsuno2012,Kimura2013}. In fact,  recent accumulation of
experimental data for the low-lying states and their electric transitions
\cite{Mijatovic2018,Momiyama2020,Longfellow2020} are revealing
the onset of the ground state deformation and the quenching of the $N=28$ shell gap in neutron-rich
Mg, Si, S and Ar isotopes. Therefore, it is important to theoretically investigate the deformation
of each isotopes and provide an insight to the mechanism behind it.

For this purpose, we apply the theoretical framework of antisymmetrized molecular dynamics
(AMD)~\cite{Kanada-Enyo2003,Kanada-Enyo2012,Kimura2016} to the neutron-rich Mg, Si, S and Ar
isotopes.  We present their energy surfaces and single-particle levels as functions of the
quadrupole deformation parameters. We also provide the excitation spectra and electric properties of
the low-lying states to compare with the observed data. It will be shown that the interplay between
the proton and neutron shell effects strongly affects the shape of the $N=28$ isotones and induces
the $\gamma$ deformation of the $N=26$ and 30 isotones. The analysis of the neutron occupation
number shows that the erosion of the $N=28$ magicity gradually occurs and there is no definite
boundaries in nuclear chart where the magicity of $N=28$ is lost. It is also found that there are
two different patterns in the  $N=28$ magicity loss.  

This paper is organized as follows. In the next section, we briefly explain the framework of AMD. In
the section~\ref{sec:result}, we present the numerical results for $N=26$, 28 and 30 isotones; the
energy surfaces, single-particle levels, neutron occupation numbers, spectra and electric
properties. Based on these numerical data, we will discuss the factors that determine the shape of
each nucleus and the relationship to the loss of the neutron magic number. Final section summarizes
this work.

\section{Theoretical framework}
The $A$-body Hamiltonian used in this study is given as,
\begin{equation}
\displaystyle \hat{H}={\sum_{i}^{A} \hat{t}_{i}}-\hat{t}_\mathrm{cm}
+\frac{1}{2}{\sum_{ij}^{A} \hat{ v}_{ij}^{\mathrm{NN}}}
+\frac{1}{2}{\sum_{ij \in\mathrm{proton}}^{Z} \hat{v}_{ij}^{\mathrm{C}}},
\end{equation}
where the Gogny D1S density functional~\cite{Berger1991} is employed as an effective nucleon-nucleon
interaction $\hat{v}_{ij}^{\mathrm{NN}}$ and  the Coulomb interaction $\hat{v}_{ij}^{\mathrm{C}}$ is 
approximated by a sum of seven Gaussians.  The center-of-mass kinetic energy $\hat{t}_\mathrm{cm}$
is exactly removed.

The variational wave function is a parity-projected Slater determinant,
\begin{equation}
\Phi^{\pi}=\hat{P}^\pi \mathcal{A} \{\varphi_{1},\varphi_{2},\dots,\varphi_{A}\},
\end{equation}
where $\hat{P}^\pi$ is the parity projection operator. In this study, we investigate the low-lying
positive-parity states. The single-particle wave packet $\varphi_{i}$ is represented by a deformed
Gaussian wave packet~\cite{Kimura2004a},
\begin{align}
 \varphi_i(\bm{r}) &= \exp
 \set{-\sum_{\sigma=x,y,z} {\nu_\sigma} \left (r_{\sigma}-Z_{i\sigma} \right)^2}
\chi_{i} \eta_{i},\\
\chi_{i}&=a_{i}\chi_\uparrow+b_{i}\chi_\downarrow, \quad
 \eta_{i}= \set{\mathrm{proton\ or\ neutron}}.
\end{align}
The variational parameters are the width $(\nu_{x}, \nu_{y}, \nu_{z})$ and the centroids $\bm Z_i$ of 
Gaussian wave packets; and spin direction $a_i$ and $b_i$. They are determined by the variation with
the constraint on the matter quadrupole deformation parameters $\beta$ and
$\gamma$~\cite{Kimura2012}. The sum of the energy and the constraint potentials, 
\begin{align}
\widetilde{E}^{\pi}(\beta , \gamma) 
&=\frac{\braket{\Phi^\pi(\beta,\gamma)|\hat{H}|\Phi^\pi(\beta,\gamma)}}
{\braket{\Phi^\pi(\beta,\gamma)|\Phi^\pi(\beta,\gamma)}} \nonumber\\ 
&+v_{\beta}(\langle \beta \rangle - \beta)^2
+v_{\gamma}(\langle \gamma \rangle - \gamma)^2.
\end{align}
is minimized to obtain the optimized wave function $\Phi^{\pi}(\beta , \gamma)$ for given values of
$\beta$ and $\gamma$. The strengths of the constraint $v_\beta$ and $v_\gamma$ are chosen
sufficiently large. In this study, the set of $(\beta , \gamma)$ is chosen on the triangular lattice
on the $\beta$-$\gamma$ plane ranging from $\beta = 0 \ \mathrm{to} \ 0.6$ with interval of $0.05$. 

After the variational calculation, we perform the angular momentum projection and the generator
coordinate method (GCM).  The optimized wave functions are projected to the eigenstate of the
angular momentum,
\begin{align}
 \Phi_{MK}^{J\pi}(\beta_{i},\gamma_{i}) = \frac{2J+1}{8\pi^2}
 \int \mathrm{d}\Omega D_{MK}^{J \ast}(\Omega)\hat{R}(\Omega)
 \Phi^{\pi}(\beta_{i}, \gamma_{i}),\label{eq:basis}
\end{align}
where $D^{J}_{MK}(\Omega)$ and $R(\Omega)$ represent the Wigner's D function and rotation operator.
The projected wave functions are superposed employing $\beta$ and $\gamma$ as the generator
coordinates,  
\begin{align}
 \Psi_{M\alpha}^{J\pi}
 =\sum_{iK}  g_{iK\alpha} \Phi_{MK}^{J\pi} (\beta_{i} , \gamma_{i}),\label{eq:gcmwf}
\end{align}
where the coefficients $g_{iK\alpha}$ and eigenenergies $E_{\alpha}$ are obtained by
solving the Hill-Wheeler equation~\cite{Hill1953}, 
\begin{align}
 & \sum_{j K^{\prime}} H_{ i K j K^{\prime}} g_{j K^{\prime} \alpha}
 =E_{\alpha} \sum_{j K^{\prime}} N_{i K j K^{\prime}} g_{j K^{\prime} \alpha},\\
\label{eq:HWg}
 &H_{i K j K^{\prime}}
=\langle \Phi_{MK}^{J\pi} (\beta_{i} , \gamma_{i}) | \hat{H} |
\Phi_{MK^{\prime}}^{J\pi} (\beta_{j} , \gamma_{j}) \rangle,\\
 &N_{i K j K^{\prime}}
= \langle \Phi_{MK}^{J\pi}(\beta_{i} , \gamma_{i}) | 
\Phi_{MK^{\prime}}^{J\pi}(\beta_{j} , \gamma_{j}) \rangle.
\end{align}

In order to discuss the breaking of the magic number, we also calculate the single-particle
configuration of the optimized wave function by the following procedure~\cite{Dote1997}. We first
transform the single-particle wave packets $\varphi_i$ into the orthonormalized basis
$\widetilde{\varphi}_p$,    
\begin{equation}
\widetilde{\varphi}_p
=\frac{1}{\sqrt{\mu_p}} \sum_{i} c_{ip} \varphi_{i},
\end{equation}
where $\mu_p$ and $c_{ip}$ are the eigenvalues and the eigenvectors of the overlap matrix $B$,
\begin{align}
\sum_{j}B_{ij}c_{jp} = \mu_p c_{ip}, \quad 
B_{ij} = \langle \varphi_{i} | \varphi_{j} \rangle.
\end{align}
With this  basis, the single-particle Hamiltonian is defined as,
\begin{align}
h_{pq} &= \braket{\widetilde{\varphi}_{p} | \hat{t} | \widetilde{\varphi}_{q}}
+\sum_{r} \braket{\widetilde{\varphi}_{p} \widetilde{\varphi}_{r} |
 \hat{v}^{\mathrm{NN}} + \hat{v}^{\mathrm{C}} |
 \widetilde{\varphi}_{q} \widetilde{\varphi}_{r} - \widetilde{\varphi}_{r} \widetilde{\varphi}_{q}
 } \nonumber\\ 
&+ \frac{1}{2} \sum_{r, s} \braket{ \widetilde{\varphi}_{r} \widetilde{\varphi}_{s} 
| \widetilde{\varphi}_{p}^{\ast} \widetilde{\varphi}_{q} \frac{\delta \hat{v}^{\mathrm{NN}}}{\delta \rho} |
\widetilde{\varphi}_{r} \widetilde{\varphi}_{s} - \widetilde{\varphi}_{s} \widetilde{\varphi}_{r}}.
\end{align} 
The eigenvalues and eigenvectors of $h_{pq}$ give the single-particle energies $\epsilon_p$ and
wave functions $\phi_\alpha$, 
\begin{align}
&\sum_{q} h_{pq} f_{q \alpha}
= \epsilon_{\alpha} f_{p \alpha},\\
&\phi_{\alpha}
=\sum_{p} f_{p \alpha} \tilde{\varphi}_{p}
=\sum_{i} \left( \sum_{p} c_{ip} \frac{1}{\sqrt{\mu_p}} f_{p \alpha} \right)  \varphi_{i}.
 \label{eq:spe}
\end{align}

\section{results and discussion}\label{sec:result}
\subsection{ground state deformation, Fermi levels and $\bm{N=28}$ magicity}
\begin{figure*}[ht]
\includegraphics[width=1.0\hsize]{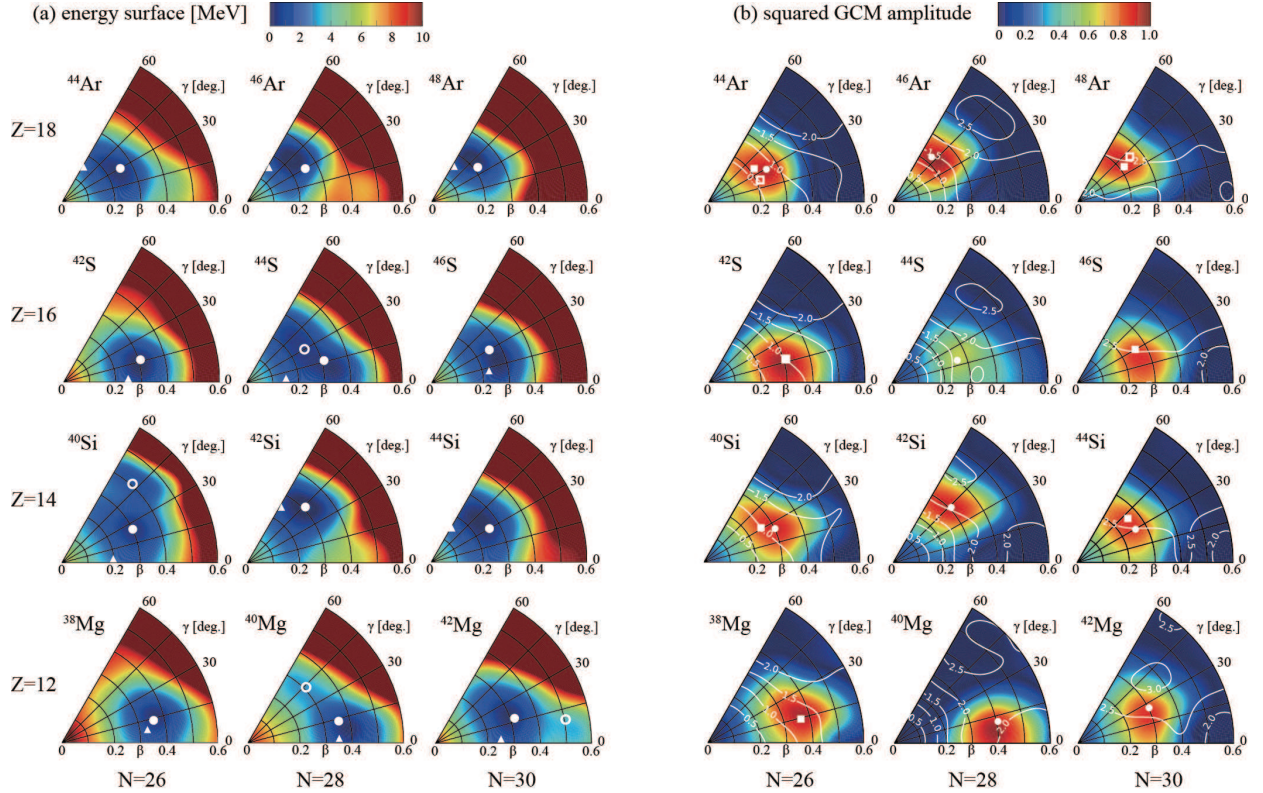}
\caption{(a) Energy surfaces of the $J^{\pi}=0^{+}$ states as functions of the quadrupole
 deformation parameters $\beta$ and $\gamma$. The filled (open) circles show the global (local)
 minima. Color plots show the energies relative to the global minima. The triangles show the energy
 minima before the angular momentum projection. (b) Squared GCM amplitude (color plots) and
 occupation number of the $1p$ orbit (contour lines) of the ground state. The filled circles and
 squares show the maxima of the squared GCM amplitudes for the ground and $2^+_1$ states,
 while open squares show the maximum for the $2^+_2$ state of $N=26$ and 30 isotones.}  
\label{fig:surface}
\end{figure*}
\begin{figure*}[ht]
\centering
\includegraphics[width=1.0\hsize]{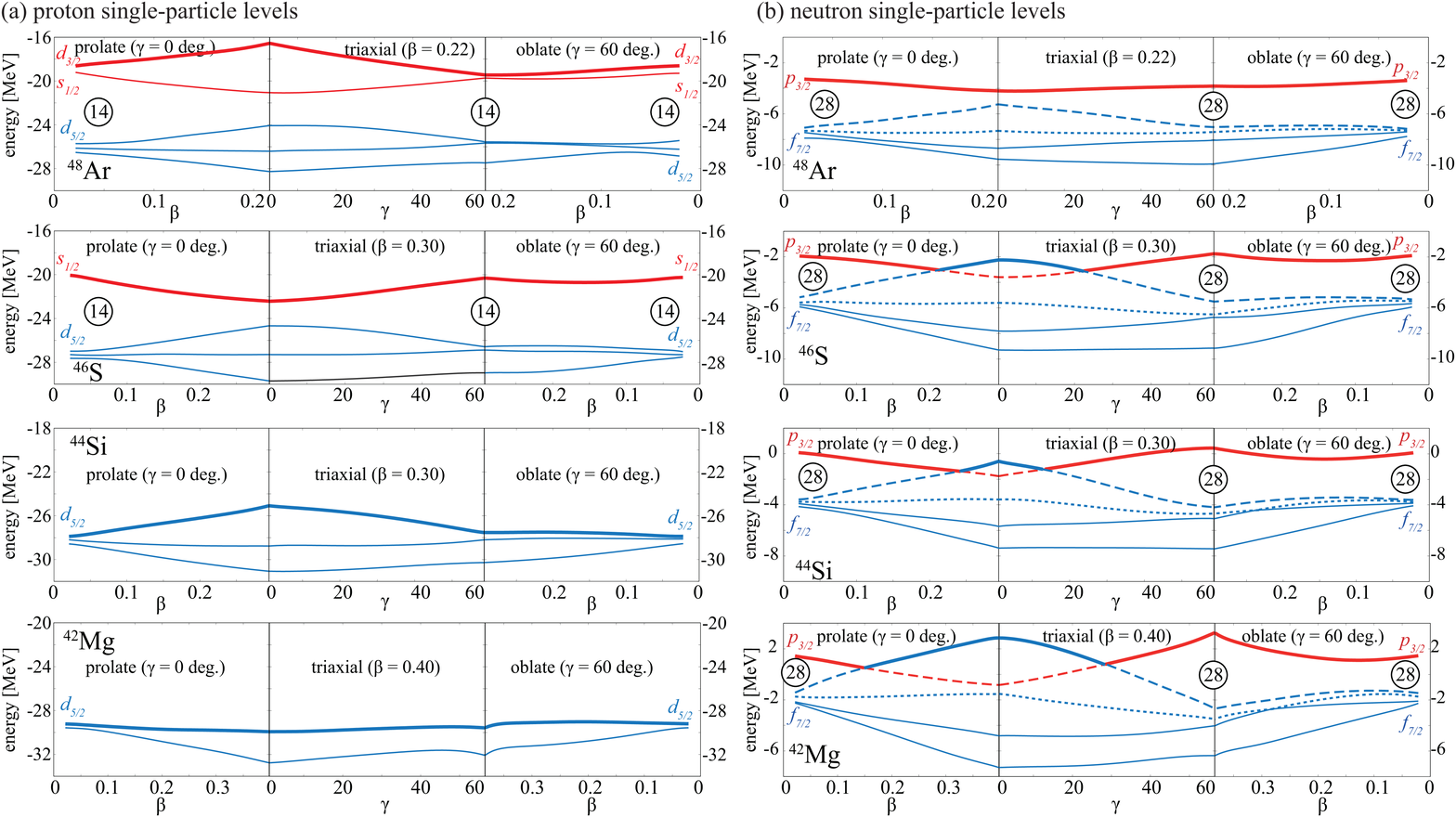}
\caption{(a) Proton single particle levels of the $N=30$ isotones as functions of the quadrupole
 deformation parameters. Left (right) panels describe prolate (oblate) deformation, while the middle
 panels describe triaxial deformation. Bold lines show the highest occupied orbits. (b) Same with
 the panels (a), but for neutrons. Bold, dashed and dotted lines show the highest occupied orbits of
 the $N=26$, 28 and 30 isotones.}  
\label{Fig:spe}
\end{figure*}

Figure~\ref{fig:surface} (a) shows the energy surfaces of $N=26$, 28 and 30 isotones as functions of the
deformation parameters $\beta$ and $\gamma$  obtained by the angular momentum projection to
$J^{\pi}=0^{+}$. It shows that all nuclei including the $N=28$ isotones have deformed energy minima
(filled circles in Fig.~\ref{fig:surface} (a)) whose deformation parameter $\beta$ are larger than 0.2.
This implies that the magicity of the neutron number $N=28$ is lost in this mass region.  Furthermore,
the energy surfaces are quite soft against $\gamma$ deformation and most of the energy minima are
triaxially deformed. For example, the energy minimum of $^{44}\mathrm{Ar}$ is at 
$(\beta, \gamma)=(0.26, 30^{\circ})$ and triaxially deformed, but its energy is very close to the
prolate and oblate deformed states with  $(\beta, \gamma)=(0.26, 0^{\circ})$ and $(0.26, 60^{\circ})$
which are only 1.5 and 1.6 MeV above the triaxial state, respectively.  We note that the angular
momentum projection is  essentially important to describe triaxial deformation. Before the
projection, all nuclei except for $^{38}\mathrm{Mg}$ and $^{46}\mathrm{S}$ have axially deformed
minima indicated by triangles in the figure,  but after the projection, triaxially deformed states
gain larger binding energy and become the ground state.
We also note that $^{40}\mathrm{Mg}$, $^{42}\mathrm{Mg}$, $^{40}\mathrm{Si}$ and $^{44}\mathrm{S}$
have low-lying local energy minima (open circles in Fig. \ref{fig:surface} (a)) at 2.8, 2.8, 1.2 and
0.19 MeV above the global minima, which generate the low-lying $0^{+}_{2}$ states.

In order to evaluate the deformation of individual nuclei in detail, we calculate the squared GCM
amplitude which is the overlap between a basis wave function (Eq.~(\ref{eq:basis})) with the
deformation parameters $(\beta, \gamma)$ and a GCM wave function (Eq.~(\ref{eq:gcmwf})),   
\begin{align}
 O_{K}^{J\pi}(\beta,\gamma)=|\braket{\Phi_{MK}^{J\pi}(\beta,\gamma)|\Psi^{J\pi}_{M\alpha}}|^2.
\end{align}
Larger amplitude means larger probability for corresponding values of the deformation parameters
($\beta$, $\gamma$), and its maximum may be regarded as the most probable intrinsic shape. The
calculated amplitudes for the ground states are shown in Fig.~\ref{fig:surface} (b). They show that
the ground states of almost all nuclei largely overlap with triaxially deformed shape with $15^\circ
< \gamma < 45^\circ$, and their maxima (filled circles) are close to the minima of the energy
surfaces shown in Fig.~\ref{fig:surface} (a). Because of triaxial deformation, the $N=26$ and 30
isotones have the non-yrast $K^{\pi} = 2^{+}$ bands built on the $2^{+}_{2}$ states as discussed
later. Figure~\ref{fig:surface} also show a trend that the Ar and Si  isotopes favor the oblate
shape ($\gamma > 30^\circ$), while S and Mg isotopes are prolate shaped ($\gamma < 30^\circ$). 

To understand the origin of the shape of each nuclei, Fig.~\ref{Fig:spe} shows the proton and
neutron single-particle orbits of the $N=30$ isotones as functions of the quadrupole deformation 
parameters. We note that $N=26$ and 28 isotones have qualitatively the same structure of 
single-particle orbits. First, we focus on the proton Fermi level (highest occupied orbit) shown by
bold lines in panel~(a). In Ar and Si, its energy is lowered by oblate deformation, which explains
the reason why these isotopes tend to manifest oblate deformation. On the other hand, in S and Mg,
the Fermi level is lowered by prolate deformation or almost flat. Thus, the proton Fermi level
causes the general deformation trend of each isotope. 

The neutron single-particle orbits show the different dependence on deformation. The neutron
Fermi level of the $N=30$ isotones (bold lines in panel~(b)) is almost constant in Ar, but as proton
number decreases, it starts to show deformation dependence. This partly owes to the quench of $N=28$ shell
gap and resultant level inversion. They induce many-body correlation leading to the rearrangement of
the single-particle levels in a self-consistent manner. Note that the spherical $N=28$ shell-gap
gets smaller as the proton number decreases. It is 4.0, 3.5, 3.4 and 2.8 MeV for $^{48}\mathrm{Ar}$,
$^{46}\mathrm{S}$, $^{44}\mathrm{Si}$ and $^{42}\mathrm{Mg}$, respectively.  Consequently,
$^{42}\mathrm{Mg}$ is strongly suffered from the shell quenching and results in the bumpy behavior
of the Fermi level. The Fermi levels of the $N=28$ isotones (dashed lines in pales~(b)) more clearly
depend on deformation and tend to favor oblate deformation except for Mg. Similarly to the $N=30$
isotones, the deformation dependence gets stronger as proton number decreases.  In contrast to
$N=28$ and 30 isotones, the Fermi levels of $N=26$ isotones show moderate dependence (dotted lines
in panel~(b)) as they are well below the $N=28$ shell gap and more deeply bound.  

The interplay between the proton and neutron shell effects affects the shape of $N=28$ isotones. Let
us explain it with a couple of examples. In the case of $^{42}{\rm Si}$, both proton and neutron
Fermi levels favor the oblate shape and coherently induce strong oblate deformation. Similarly, 
the proton and neutron shell effects cooperatively bring about oblate deformation to $^{46}{\rm Ar}$
and prolate deformation to $^{40}{\rm Mg}$. To the contrary, the proton and neutron shell effect
act in the opposite way for $^{44}{\rm S}$; proton's Fermi level favors  prolate deformation whereas the
neutron's one favors oblate deformation. As a result, the energy surface of $^{44}{\rm S}$ is
double-well shaped with two minima in prolate and oblate sides, and the many body correlation
induced on top it makes the low-lying spectrum of $^{44}{\rm S}$ rather complex compared to
neighboring nuclei. Indeed, many low-lying non-yrast states have been
observed~\cite{Santiago-Gonzalez2011,Parker2017} in this nucleus and their unique nature has been
discussed~\cite{Rodriguez2011,Utsuno2015}.  

As mentioned above, most of the $N=26$ and 30 isotones have flat neutron Fermi levels insensitive to 
deformation, so their energy surfaces are also flat to deformation. This activates the
degree-of-freedom of $\gamma$ deformation and generates the low-lying $K^\pi=2^+$ bands to be
discussed in the next section. An exceptional case is $^{42}{\rm Mg}$ in which neutron Fermi level
shows bumpy behavior due to the quenched $N=28$ shell gap and level inversion. At $\gamma=30^\circ$,
the neutron Fermi energy is low and high single-particle level density induces pairing correlations
to gain larger binding energy. Consequently, $^{42}{\rm Mg}$ has a well-developed triaxially
deformed energy minimum. Another point mentioned for this nucleus is that it is not bound by the
Gogny D1S density functional used in this study. This result contradict to the recent shell model
study~\cite{Tsunoda2020}, and hence, the experimental information on the binding of this nucleus will
give us a deeper understanding of nuclear force. 

The last issue to mention in Fig. 2 is the loss of the $N=28$ magicity. In the prolate deformed
region, the level inversion takes place between the neutron single-particle levels which originate
in the spherical $0f_{7/2}$ and $1p_{3/2}$. In the case of the $N=30$ isotones, it occurs at
$\beta$=0.22,  0.24 and 0.15 in $^{46}\mathrm{S}$, $^{44}\mathrm{Si}$ and $^{42}\mathrm{Mg}$,
respectively.  On the other hand, the $N=28$ shell gap is kept large in the oblate deformed region,
and hence, we are tempted to conclude that the magicity of the neutron number 28 is robust in the
oblate deformed nuclei such as $^{42}\mathrm{Si}$. However, as we show below, the $N=28$ magicity is
also lost in oblate and triaxial deformed nuclei even though there is no explicit inversion of the
single-particle levels. To elucidate this, we consider the multipole decomposition of the
single-particle orbits (Eq.~(\ref{eq:spe})) as,
\begin{align}
\phi_i(\bm{r})
= \sum_{jlj_z} \phi_{i;jlj_z}(r) [Y_{l}(\hat r) \times\chi_{1/2}]_{jj_z}.
\end{align}
The squared integral of the $l=1$ components ($p_{3/2}$ and $p_{1/2}$) gives us an estimate of the
occupation number of the neutron $1p$-orbit. Assuming the complete filling of the $0p$-orbit, the
neutron occupation number of the $1p$-orbit is obtained as,
\begin{align}
N_{1p}= \sum_{i=1}^{N} \sum_{jj_z} \int_0^\infty |r\phi_{i;j1j_z}(r)|^2dr  - 6. \label{eq:oc}
\end{align}

The contours in Fig.~\ref{fig:surface} (b) show the occupation numbers of the $1p$ orbits as
functions of the deformation parameters. It must be noted that the occupation number becomes large
not only in the prolate deformed region but also in the oblate deformed region even though there is
no single-particle level inversion. This indicates that the deformation causes the strong mixing of
the $l=3$ and 1 components in the single-particle levels close to the Fermi surface to increase the
$p$-wave occupation probability. Consequently, there are two different types in the loss of
the $N=28$ magicity: The magicity is lost due the single-particle level inversion in the prolate
deformed nuclei, while it is lost without the inversion in the oblate deformed nuclei. 

\begin{figure}[ht]
\centering
\includegraphics[width=0.4\hsize]{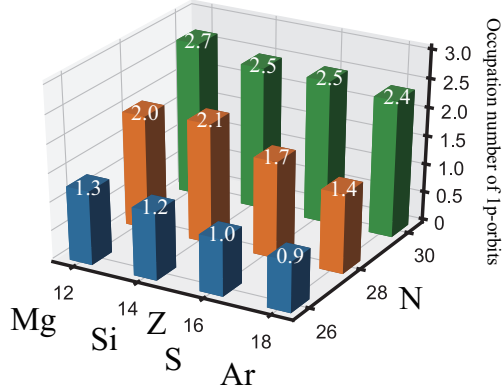}
 \caption{Occupation number of the neutron $1p$ orbits at the position where the GCM amplitude of
 the ground state is maximum.} \label{fig:oc}
\end{figure}

The maximum of the GCM amplitude of $N=28$ isotones ($^{42}{\rm Ar}$,$^{44}{\rm S}$, $^{42}{\rm Si}$
and $^{40}{\rm Mg}$) are located at the regions where the occupation number is close to or larger
than 2. Therefore, their ground states are dominated by the $2p2h$ configurations because of the
loss of the $N=28$ magicity. To illustrate the landscape of the magicity loss, Fig.~\ref{fig:oc}
shows the occupation number of the $1p$ orbits at the maximum of the GCM amplitudes. It shows that
the occupation number gradually and continuously increases as proton number decreases and 
neutron number increases. Therefore, it seems that there is no clear boundary in nuclear chart where
the  $N=28$ magicity is lost. This feature is different from the case of the island of inversion
which seems to have distinct boundary at $N=19$ and
$Z=13$~\cite{Warburton1990,Han2017,Shimada2012,Heylen2016}.

\subsection{Low-lying spectrum and triaxial deformation}
\begin{figure*}[bht]
\centering
\includegraphics[width=0.8\hsize]{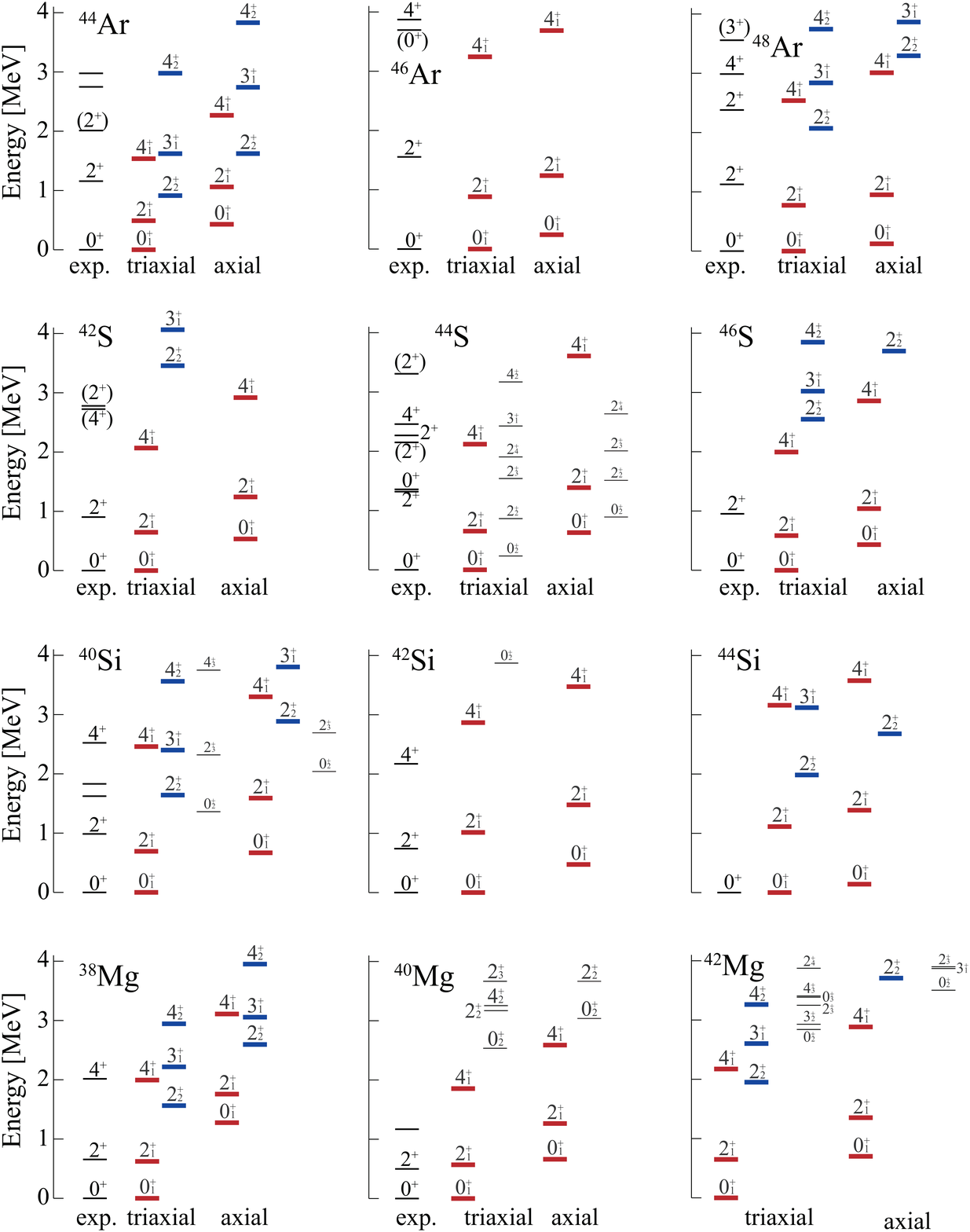}
\caption{Low-lying positive-parity spectra of the $N=26$, 28 and 30 isotones up to $J^{\pi}=4^{+}$
 compared to the experimental data. ``triaxial'' denotes the results of the GCM
 calculations which allow $\gamma$ deformation, while ``axial'' denotes that limited to axial
 symmetric shape ($\gamma=0^\circ$ or $60^\circ$). }\label{fig:spectrum} 
\end{figure*}
\begin{table*}
\caption{The calculated and observed $B(E2)$ strengths in the unit of $e^{2} \mathrm{fm}^4$ and
 electric quadrupole moments in the unit of $e \mathrm{fm}^2$.  ``triaxial'' denotes the results
 obtained by the GCM calculations  which allow $\gamma$ deformation, while ``axial'' denotes that
 restricted to axial symmetric shape.}   
\label{tab:E2}
\begin{center}
\begin{tabular}{cc|cccccccccccc} \toprule
 &  & 
$^{38}\mathrm{Mg}$ & $^{40}\mathrm{Si}$ & $^{42}\mathrm{S}$ & $^{44}\mathrm{Ar}$ &
$^{40}\mathrm{Mg}$ & $^{42}\mathrm{Si}$ & $^{44}\mathrm{S}$ & $^{46}\mathrm{Ar}$ &
$^{42}\mathrm{Mg}$ & $^{44}\mathrm{Si}$ & $^{46}\mathrm{S}$ & $^{48}\mathrm{Ar}$ \\ \hline
$B(E2; 0_{1}^{+} \to 2^{+}_{1})$ &triaxial & 440 & 209 & 406 & 231 & 486 & 365 & 383 & 334 & 341 & 221 & 359 & 346 \\
                                      & axial    & 357 & 162 & 416 & 286 & 474 & 356 & 145 & 326 & 290 &194 & 326 & 321 \\
                                      & exp.      &  &  & $397^{+63}_{-63}$ & $378_{-55}^{+34}$ &  &  & $314^{+88}_{-88}$ &  $216^{+22}_{-22}$ \cite{Calinescu2016} &  &  &  & $346^{+55}_{-55}$ \\
                                      &             &  &  &             &                            &  &  &             & $570^{+335}_{-160}$ \cite{Mengoni2010} &  &  &  &             \\ \hline
$B(E2; 0_{1}^{+} \to 2^{+}_{2})$ & triaxial & 18 & 90 & 58 & 88 & 0.03 &  & 8.4 &  & 26 & 15 & 89 & 7.1 \\
                                       & axial    & 9.8 & 40 & 14 & 6.0 & 0.06 &  & 10 &  & 4.1 & 7.9 & 36 & 1.4 \\
                                      & exp.      &  &  &  & $23^{+2}_{-2}$ &  &  &  &  &   &  &  &  \\ \hline
$B(E2; 2^{+}_{1} \to 2^{+}_{2})$ & triaxial & 11 & 45 & 14 & 72 & 0.02 &  & 2.5 &  & 11 & 68 & 14 & 17 \\
                                           & axial     & 0.2 & 16 & 3.5 & 42 & 0.07 &  & 14 &  & 9.6 & 53 & 7.3 & 2.2 \\
                                           & exp.      &  &  &  & $680_{-90}^{+150}$ &  &  &  &  &  &  &  &  \\ \hline
$Q(2^{+}_{1})$  & triaxial & -18.5 & -10.3 & -18.4 & -7.0 & -20.1 & 17.3 & -18.0 & 16.8 & -16.3 &  2.1 & -18.1 & 15.8 \\          
                     & axial     & -17.4 & -6.4 & -18.7 & 11.5 & -19.6 & 17.6 & -12.1 & 16.5 & -15.0 & 5.3 & -17.0 & 16.4 \\   
                     & exp.      &  &  &  & -$8.3^{+3}_{-3}$ &  &  &  &  &  &  &  &  \\ \hline           
\end{tabular}
\end{center}
\end{table*} 

Here, we discuss how quadrupole deformation, in particular the $\gamma$ deformation,
influences to the excitation spectrum. Figure~\ref{fig:spectrum} shows the energy spectra obtained by
the GCM calculations which are labeled as ``triaxial''. Compared to a closed-shell nucleus
$^{48}\mathrm{Ca}$ which has the $2_{1}^{+}$ and $4_{1}^{+}$ states at 3.8 MeV and 4.5
MeV~\cite{Burrows2006},  all of the calculated Mg, Si, S and Ar isotopes have low-lying $2_{1}^{+}$
and $4_{1}^{+}$ states, and the energy ratios $E(4_{1}^{+})/E(2_{1}^{+})$ are close to 3.3,
indicating their rotational nature.  These characteristics are consistent with the observed spectra,
although the calculation overestimates the moments-of-inertia of several nuclei. Because of the 
deformation, the $E2$ transition strengths ($0_1^+ \to 2^+_1$) listed in table~\ref{tab:E2}
are stronger than 200 $e^2 \mathrm{fm}^4$ in all nuclei which are more than twice as large as that
of $^{48}\mathrm{Ca}$~\cite{Hartmann2002a}. All these results are in accordance with the loss of the
$N=28$ magicity.  

In addition to the rotational ground band, all the $N=26$ and 30 isotones have the low-lying
non-yrast $K^\pi=2^+$ bands built on the $2^+_2$ states because of their pronounced triaxial
deformation. Note that the maximum GCM amplitudes of the $2^+_1$ and $2^+_2$ states
(squares and triangles in Fig.~\ref{fig:surface} (b)) are located at almost the same position
showing that the intrinsic shape of the ground band and $K^\pi=2^+$ band are similar to each other.  
Experimentally, the candidates of the $2^+_2$ state have been observed in $^{40}\mathrm{Si}$,
$^{42}\mathrm{S}$, $^{44}\mathrm{Ar}$ and $^{48}\mathrm{Ar}$. The small excitation
energies of the $3^+$ state as a member of the $K^\pi = 2^+$ band is another signature of triaxial
deformation.  The candidates such $3^+$ state have also been observed at 4.2 MeV in
$^{42}\mathrm{S}$ and 3.3 MeV in  $^{48}\mathrm{Ar}$.

To investigate how the degrees of $\gamma$ deformation affect the low-lying spectroscopy, we
have performed additional GCM calculations which are restricted to axial deformation. The obtained
spectra are labeled as ``axial'' in Fig.~\ref{fig:spectrum}.  It is found that the energy of the
non-yrast bands is sensitive to the degrees of $\gamma$ deformation.  Namely, the $K^{\pi} = 2^{+}$
bands of the $N=26$ and 30 isotones calculated by the axial GCM are much less bound than the
triaxial results. On the other hand, the energies of the ground states are not strongly affected by
the degree of triaxial deformation except for $^{38}\mathrm{Mg}$. The axial and triaxial GCM
calculations also yielded largely different inter-band transition strengths between the ground and
$K^\pi=2^+$ bands as listed in Table~\ref{tab:E2}. The $0_1^+ \to 2^+_2$ and $2^+_1 \to 2^+_2$
transitions obtained by the axial GCM are much smaller than those obtained by the triaxial GCM except
for $^{44}\mathrm{Si}$. In short, the degree-of-freedom of $\gamma$ deformation is mostly
reflected to the properties of the non-yrast $K^\pi=2^+$ band. We also remark that the triaxial GCM 
reproduces the quadrupole moment of the $2^+_1$ states of $^{44}\mathrm{Ar}$ while the axial GCM does
not. Thus, the electric quadrupole transitions and moments are sensitive probes for the triaxial
deformation. In particular, the properties of the non-yrast $K^{\pi} = 2^+$ band is important 
to identify the triaxial deformation of the $N=26$ and 30 isotones.

\section{summary}\label{sec:summary}
In this work, we aimed to investigate the nature and cause of the nuclear deformations and its
relation to the loss of the neutron magicity $N=28$ in vicinity of  $^{42}{\rm Si}$. For this
purpose, we have employed a theoretical framework of AMD combined with the Gogny D1S density
functional to calculate the neutron-rich $N=26$, 28 and 30 isotones. We have presented the energy
surfaces, GCM amplitudes and single-particle levels as functions of the quadrupole deformation
parameters. We have also shown the spectra and electric properties of the low-lying states,
which qualitatively agree with the observed data and predict the existence of the many
non-yrast states.   

The analysis of the energy surfaces and the single particle levels has revealed that the interplay
between the proton and neutron shell effects strongly affects shape of the $N=28$ isotones. In the
case of $^{46}{\rm Ar}$, $^{42}{\rm Si}$ and $^{40}{\rm Mg}$, they cooperatively act and induce
oblate or prolate deformation. On the other hand, they act in the opposite way and make the
double-well shaped energy surface of $^{44}{\rm S}$. Consequently, the low-lying spectrum of
$^{44}{\rm S}$ is relatively complicated and many a couple of $2^+$ states coexist within small
excitation energy. In the case of the $N=26$ and 30 isotones, except for $^{42}{\rm Mg}$, the
neutron Fermi level is insensitive to deformation. As a result, the energy surfaces become quite
soft against $\gamma$ deformation, and many of the $N=26$ and 30 isotones have the triaxially
deformed ground states. We have pointed out that the triaxial deformation of the $N=26$ and 30
isotones is mostly reflected in the properties of the non-yrast $K^\pi=2^+$ bands, especially in
their inter-band $E2$ transition strengths.  

Finally, we note that the erosion of the $N=28$ magicity gradually occurs as seen in the
occupation number of the neutron $1p$ orbits. Therefore, there is no definite boundaries in 
nuclear chart where the magicity of $N=28$ is lost. This is unlike to the case of the magic number
$N=20$ (island of inversion). We also remark that there are two different patterns of the 
$N=28$ magicity loss. In the case of the prolate deformed nuclei such as $^{42}{\rm Mg}$, the
magicity is lost by the inversion of single-particle levels, while it is lost without the inversion
in the case of the oblate deformed nuclei such as $^{42}{\rm Si}$. In the forthcoming paper, we will
revisit this issue and discuss how to distinguish these two patterns experimentally. 

\begin{acknowledgements}
One of the author (M.K.) acknowledges that this work was supported by the JSPS KAKENHI Grant
No. 19K03859 and by the COREnet program at RCNP Osaka University. Part of the numerical calculations
were performed using Oakforest-PACS at the Center for Computational Sciences in the University of
Tsukuba.    
\end{acknowledgements}

\bibliography{n28}

\end{document}